\documentclass[aps,prx,twocolumn,notitlepage,showkeys,showpacs,superscriptaddress,longbibliography,footinbib, preprintnumbers]{revtex4-2}
\usepackage{epsf}
\usepackage{here,times}
\usepackage[colorlinks=true,citecolor=blue,linkcolor=blue]{hyperref} 
\usepackage{slashed}
\usepackage{amsthm}
\usepackage{thmtools}
\usepackage{blindtext}
\usepackage{bm, bbm}
\usepackage{mathrsfs}
\usepackage{amsfonts, mathtools, amsmath, amssymb, dsfont}
\usepackage{esvect}
\usepackage{enumerate,dcolumn,multirow, enumitem}
\usepackage{longtable}
\usepackage{array}
\usepackage{bbding}
\usepackage[normalem]{ulem}
\usepackage{mathrsfs}
\usepackage{calligra}
\usepackage{cancel}
\usepackage{orcidlink}

\usepackage{tikz}
\usetikzlibrary{matrix,positioning, decorations.pathreplacing,calligraphy,decorations.markings, knots}

\definecolor{tensorcolor}{rgb}{1., 1., 1.}
\definecolor{btensorcolor}{rgb}{0.65,0.50,0.69}

\newcommand{\calE}{\mathcal{E}}

\newcommand{\transpose}{{\mathsf{T}}}

\newcommand{\Tr}{\mathrm{Tr}}

\begin{document}
\title{Toward bootstrapping tensor-network contractions}

\author{Seishiro Ono\orcidlink{0000-0002-4908-2612}}
\affiliation{Institute for Solid State Physics, University of Tokyo, Kashiwa 277-8581, Japan}
\affiliation{Department of Physics, The Hong Kong University of Science and Technology, Clear Water Bay, Hong Kong, China}
\affiliation{RIKEN Center for Interdisciplinary Theoretical and Mathematical Sciences (iTHEMS), RIKEN, Wako 351-0198, Japan}

\author{Yanbai Zhang}
\affiliation{Department of Physics, The Hong Kong University of Science and Technology, Clear Water Bay, Hong Kong, China}
\affiliation{Center for Theoretical Condensed Matter Physics, The Hong Kong University of Science and Technology, Clear Water Bay, Hong Kong, China}

\author{Hoi Chun Po}
\email{hcpo@ust.hk}
\affiliation{Department of Physics, The Hong Kong University of Science and Technology, Clear Water Bay, Hong Kong, China}
\affiliation{Center for Theoretical Condensed Matter Physics, The Hong Kong University of Science and Technology, Clear Water Bay, Hong Kong, China}

\preprint{RIKEN-iTHEMS-Report-26}

\begin{abstract}
    Accurate contraction of tensor networks beyond one dimension is essential in various fields including quantum many-body physics.
    Existing approaches typically rely on approximate contraction schemes and do not provide certified error bars.
    We introduce a numerical bootstrap framework which casts the problem of tensor-network contractions into a convex optimization problem, thereby yielding certified lower and upper bounds on expectation values of physical observables.
    As a proof-of-principle, we construct such constraints explicitly for translationally invariant matrix product states and demonstrate that, assuming a canonical form, second-order-cone relaxation can provide tight bounds on the contraction result. We further demonstrate that when the requirement on canonical form is lifted, a more general semidefinite-programming approach could yield similar tight bounds at higher but still polynomial computational cost.
    Our work suggests numerical bootstrap could be a possible way forward for the rigorous contractions of tensor networks.
\end{abstract}
\maketitle

\section{Introduction}
Tensor networks have become indispensable across quantum many-body physics, statistical mechanics, and machine learning~\cite{TN_review1,TN_review2, TN_review3,TN_ML1,TN_ML2,TN_ML3,TN_ML4,TN_ML5}, as they provide compact representations of high-dimensional objects including quantum many-body states and operators.
Their structure naturally encodes locality and correlations, enabling highly efficient numerical algorithms.
In one dimension, matrix product states~(MPSs) underpin state-of-the-art methods for computing ground states~\cite{White_DMRG,DMRG1}.

Beyond MPSs, higher-dimensional tensor networks, such as projected entangled pair states~(PEPS)~\cite{boundary-MPS1}, provide a natural framework for describing strongly correlated systems and their dynamics.
They have been applied to, for example, quantum spin liquids in frustrated magnets~\cite{Spin-Liquid1,Spin-Liquid2,Spin-Liquid3,Spin-Liquid4,Spin-Liquid5,Spin-Liquid6,Spin-Liquid7,Spin-Liquid8}, critical behavior of statistical-mechanical models~\cite{Baxter_Transfer, Gu-Wen-Entanglement-filtering, Classical_Ueda1,Classical_Ueda2}, and out-of-equilibrium real-time evolution~\cite{TEBD,TDVP1,TDVP2}.
In these settings, the relevant tensor networks typically contain closed loops, and local expectation values are obtained by contracting such networks.

Tensor-network contractions in higher-dimensional networks are fundamentally harder than those in MPSs due to the presence of closed loops, which render the faithful contraction of the network exponentially hard in general~\cite{PhysRevLett.98.140506}.
To avoid such prohibitive computational cost, commonly used algorithms rely on approximate contraction schemes~\cite{CTMRG1,CTMRG2,CTMRG3,boundary-MPS1, boundary-MPS2,Levin-Nave-TRG,STNR,EV-TNR,STNR,HOSVD, Loop-TNR,HOTRG2,ATRG,CTM-TRG,BW-TRG, Global_TNR_Ueda,BP1,BP2,BP3,BP4,BP5}, which can demonstrate good convergence with system sizes but typically do not provide certified error bars.

In this work, we explore an alternative approach that certifies tensor-network contraction results by providing guaranteed lower and upper bounds.
We develop a numerical bootstrap approach which derives tractable necessary conditions constraining the effective environment state surrounding the target site(s) on which physical observables are evaluated.
Unlike existing direct approaches which approximate the environment state with tenable computational resources, the bootstrap paradigm takes a dual perspective and instead focuses on necessary conditions which constraint the possible space of environment states. Importantly, the set of constraints is additive, in that constraints derived in one approach can generally be combined with those obtained in other means. In the context of higher-dimensional tensor-network contraction, the ultimate goal is to develop a systematic approach to identify and then combine effective bootstrap constraints on the environment state coming from, say, different spatial directions.

As a first step toward the stated ambitious goal, here we introduce a concrete framework for constructing such constraints for MPSs. We demonstrate numerically how the resulting bounds shrink exponentially toward the exact result in the thermodynamic limit.
The key ingredients for MPSs extend conceptually to higher dimensional problems, although making them practical remains an outstanding challenge which lies beyond the scope of the present work.

We note that semidefinite programming (SDP) relaxations of the set of physical density matrices have already been deployed to derive lower bounds on ground-state energies~\cite{QChem_boot1, QChem_boot2, Eslam_bootstrap_1, RDM_PRX} and two-sided bounds on expectation values with respect to ground states~\cite{PhysRevB.94.195143, Certificate_GS_1, Certificate_GS_2}.
As is shown below, the numerical bootstrap approach we develop differs from these existing schemes in that the constraints are not derived from the relations between physical observables; rather, they arise from the interpretation of a tensor network as a positive linear map propagating the usually arbitrary boundary conditions to the environment state surrounding the target sites. 
Furthermore, a certified contraction of tensor networks also has application beyond ground-state problems. For instance, the problems of evaluating partition functions of classical lattice models and simulating time evolutions of quantum many-body systems could both be cast as a tensor-network contraction problem~\cite{Baxter_Transfer, CTMRG1, TEBD}.

\section{Results}
\subsection{General idea}
\begin{figure}[t]
	\centering
	\includegraphics[width=0.99\columnwidth]{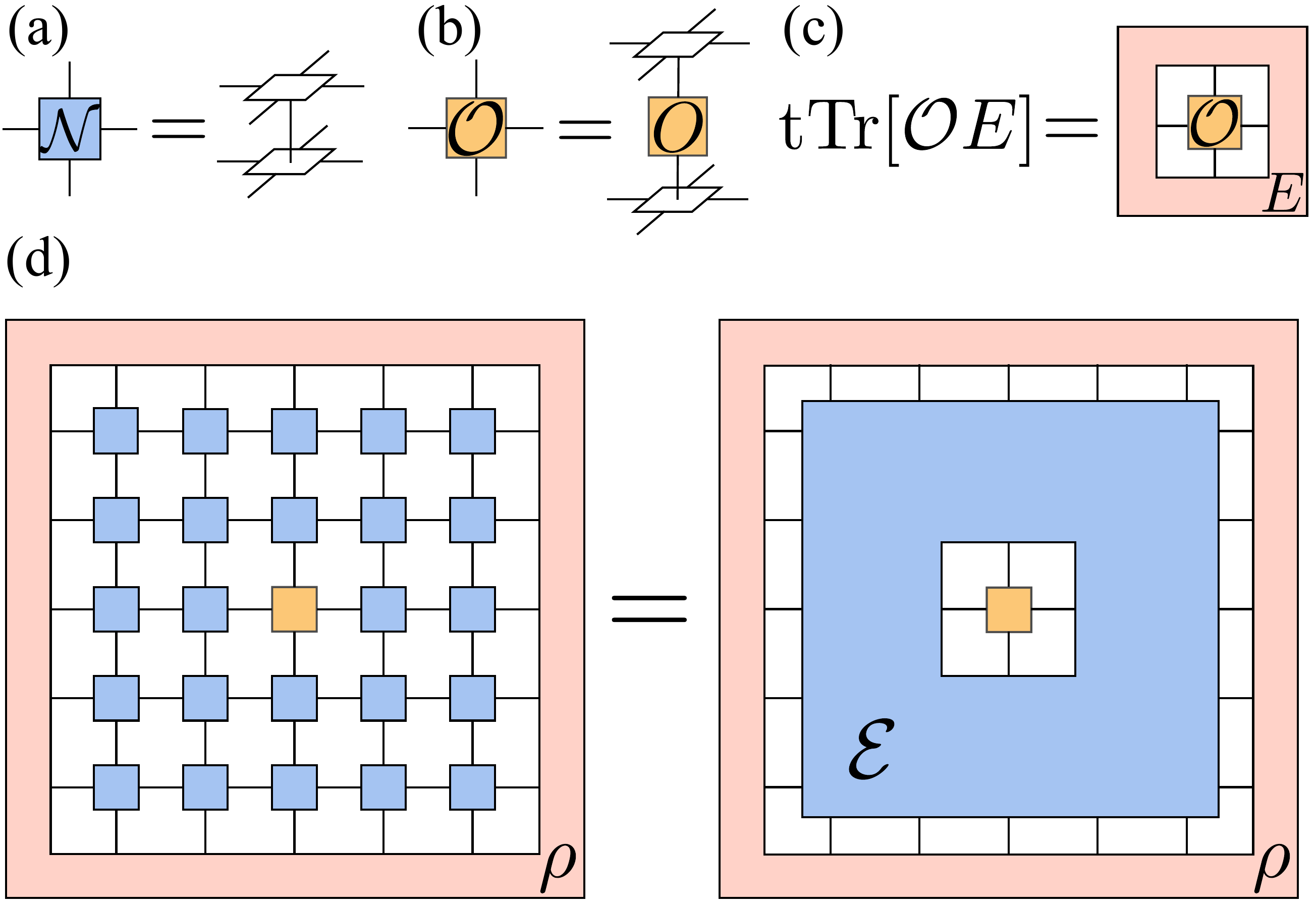}
	\caption{\label{fig:ipeps}
	Illustration of an example of two-dimensional tensor networks.
	(a) A transfer matrix defined by a local tensor of PEPS.
	(b) An operator acting on the local bond space.
	(c) A diagrammatic expression of the expectation value of a local operator $O$ with respect to a given environment $E$.
	(d) An expectation value of a local operator $O$ for a given boundary environment $\rho$. Tensors surrounding the target site form a linear map $\calE:\mathbb{M}_{\chi^{4N}}(\mathbb{C})\rightarrow \mathbb{M}_{\chi^{4}}(\mathbb{C})$.
	}
\end{figure}

We begin by describing how general tensor-network contraction problems could be viewed as an optimization problem, how the problem could be relaxed to provide certified two-sided bounds to the contraction results.
As an example, consider the expectation value of a local observable $O$ with respect to a two-dimensional translationally invariant PEPS.
The PEPS is defined by a rank-5 tensor $\{A^{s}_{lurd}\}_{l,u,r,d,s}$ with bond dimension $\chi$ and local Hilbert space dimension $D$.
The resulting two-dimensional tensor network is illustrated in Fig.~\ref{fig:ipeps}.
The constituent double-layer tensors are defined by $\mathcal{N}^{lurd}_{l'u'r'd'} = \sum_{s=0}^{D-1} A^{s}_{lurd} [A^{s}_{l'u'r'd'}]^{*}$ and $\mathcal{O}^{lurd}_{l'u'r'd'} = \sum_{s, s'=0}^{D-1} A^{s}_{lurd} O_{s's}[A^{s'}_{l'u'r'd'}]^{*}$ in Fig.~\ref{fig:ipeps}(a,b), respectively.
We focus on a single site located at the center of a finite region, for example, an $N \times N$ subregion, within the tensor network.
The surrounding part of the network can be regarded as a completely positive linear map $\calE:\mathbb{M}_{\chi^{4N}}(\mathbb{C})\rightarrow \mathbb{M}_{\chi^{4}}(\mathbb{C})$, where $\mathbb{M}_{n}(\mathbb{C})$ denotes the set of $n \times n$ complex matrices.

We first note that the evaluation of local observables in tensor networks can be formulated as an optimization problem.
Given a positive operator $\rho \in \mathbb{M}_{\chi^{4N}}(\mathbb{C})$ at the boundary of the subregion, an effective environment operator is obtained by applying the linear map $\mathcal{E}$ to $\rho$ [see Fig.~\ref{fig:ipeps}(d)].
Instead of specifying the boundary operator $\rho$, which corresponds to choosing a specific boundary condition, we only require the positive semidefiniteness of $\rho$, i.e., $\rho \succeq 0$.
Consequently, the expectation value of an observable $O$ can be rigorously bounded by the following convex programming problem:
\begin{equation}\label{eq:bare_optimization}
	\begin{split}
		\min_{\rho \succeq 0} \text{ or }\max_{\rho \succeq 0}&\ \ \mathrm{tTr}[\mathcal{O} \calE(\rho)] \\
		\text{subject to} \ \ & \mathrm{tTr}[\mathcal{N} \calE(\rho)] = 1
	\end{split}\quad,
\end{equation}
where $\mathrm{tTr}[\mathcal{O} X] = \sum_{\text{all indices}} \mathcal{O}^{lurd}_{l'u'r'd'} X^{lurd}_{l'u'r'd'}$ (see Fig.~\ref{fig:ipeps}(c)), and $\mathrm{tTr}[\mathcal{N} \calE(\rho)]=1$ originates from the normalization of states.
Since the optimization is performed over the set of all possible positive operators at the boundary, the resulting minimum and maximum values provide rigorous lower and upper bounds on the expectation value, denoted by $O_{\text{min}}^{(\text{true})}$ and $O_{\text{max}}^{(\text{true})}$, respectively. A tight two-sided bound could be obtained only if the expectation value of $\mathcal O$ is independent of the boundary condition in the thermodynamic limit.

While general, the optimization problem \eqref{eq:bare_optimization} becomes intractable for large subregions because the dimension of $\rho$ grows exponentially with $N$.
We tackle this problem in the spirit of numerical bootstrap, namely, we relax the problem by considering only necessary conditions for the propagated environment state $\mathcal E(\rho)$.
Let $E(\mathbf{x})$ be a $\chi^{4}$-dimensional Hermitian matrix parameterized by a parameter set $\mathbf{x} \in \mathbb{R}^{r}\ (r \leq \chi^8)$.
We are interested in deriving a collection of necessary conditions for $E$ to be representable as $E = \calE(\rho)$ for some $\rho \succeq 0$. The necessary conditions we derive will take the following form: (i) $\mathcal{M}(\mathbf{x}) \succeq 0$ for a linear Hermitian-matrix-valued function $\mathcal{M}$ of $\mathbf{x}$;~(ii) $f(\mathbf{x}) \geq 0$ for a scalar-valued concave function $f:\mathbb{R}^{r}\rightarrow \mathbb{R}$.
These conditions effectively define an outer convex approximation of the set of physically realizable environments.
Accordingly, we solve the following convex optimization problem:
\begin{equation}
	\label{eq:problem}
	\begin{split}
		\min_{\mathbf{x}\in \mathbb{R}^{r}} \text{ or } \max_{\mathbf{x}\in \mathbb{R}^{r}} &\ \mathrm{tTr}[\mathcal{O} E(\mathbf{x})] \\
		\text{subject to} \ & \mathrm{tTr}[\mathcal{N} E(\mathbf{x})] = 1 \\
		& f_{i}(\mathbf{x}) \geq 0\ \text{ for all } i=1, \cdots, n_{\text{SOC}}\\
		& \mathcal{M}_{j}(\mathbf{x}) \succeq 0\ \text{ for all } j=1, \cdots, n_{\text{LMI}}
	\end{split}\quad,
\end{equation}
where $n_{\text{SOC}}$ and $n_{\text{LMI}}$ denote the number of scalar-type and matrix-type constraints, respectively.
The problem \eqref{eq:problem}, which replaces the exact constraints in \eqref{eq:bare_optimization} by a collection of necessary conditions, represent an outer relaxation of the possible set of environment states.
Correspondingly, the bounds $O^{\text{(relaxed)}}_{\min/\max}$ obtained by solving problem \eqref{eq:problem} must satisfy
\begin{align}
	O^{\text{(relaxed)}}_{\min} \leq O^{\text{(true)}}_{\min} \leq O^{\text{(true)}}_{\max} \leq O^{\text{(relaxed)}}_{\max},
\end{align}
and they provide a certified bounds on the true expectation values.
The task, therefore, is to find the tractable set of necessary conditions defining problem \eqref{eq:problem}.

\subsection{Warm up: Single qubit example}
As Fig.~\ref{fig:ipeps} illustrates, in a tensor network the boundary state $\rho$ is propagated to the bonds surrounding the target sites by a positive linear map $\mathcal E$. For translation-invariant tensor networks, the map $\mathcal E$ can generally be broken down into the successive application of the same basic map defined by the local tensors.
Naturally, one expects the set of possible output states is narrowed down as the same positive map is repeatedly applied.
As a warm-up, we first demonstrate this point through the simplest example of a primitive, invertible, and completely positive trace-preserving (CPTP) map $\mathcal E$ acting on a single qubit.
For a single qubit, the normalized density matrix $\rho$ is parameterized by $\rho(\mathbf{x}) = (x_0 \sigma_0 + \tilde{\mathbf{x}}\cdot\bm{\sigma})/2$ with $x_0 = 1$, where $\sigma_0$ is an identity matrix, $\bm{\sigma}=(\sigma_1, \sigma_2, \sigma_3)$ are Pauli matrices, and $\mathbf{x} = (x_0, \tilde{\mathbf{x}}) \in \mathbb{R}^4$.
The positivity of $\rho$ implies that the parameters belong to $\mathcal{B}_0 \coloneqq \{\mathbf{x} \in \mathbb{R}^4 : \rho(\mathbf{x}) \succeq 0\ \&\ \Tr[\rho(\mathbf{x})] = 1\} = \{\mathbf{x} \in \mathbb{R}^4 : \|\tilde{\mathbf{x}}\|_{2} \leq x_0\ \&\ x_0 = 1 \} \simeq \{\tilde{\mathbf{x}} \in \mathbb{R}^3 :  \|\tilde{\mathbf{x}}\|_2 \leq 1\}$, which is the so-called \textit{Bloch ball}.
In fact, $\rho(\mathbf{x}) \succeq 0$ and $x_0 - \|\tilde{\mathbf{x}}\|_{2} \geq 0$ are instances of matrix- and scalar-type constraints, respectively.

\begin{figure}
	\centering
	\includegraphics[width=0.99\columnwidth]{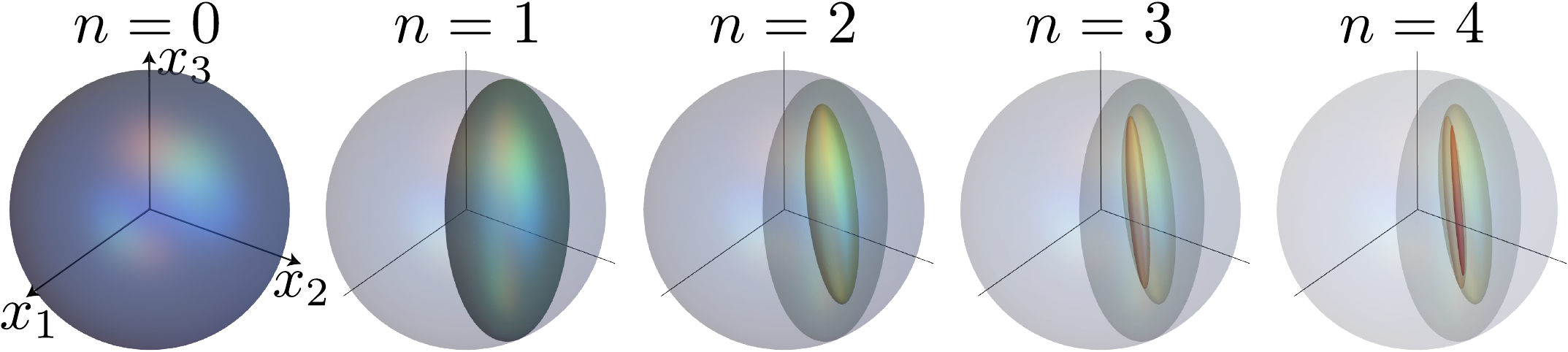}
	\caption{\label{fig:bloch}
	Parameter space $\mathcal{B}_{n}$ of density matrices for a single qubit for $n=0,1,2,3,4$.
	For comparison, $\{\mathcal{B}_{m}\}_{m=0}^{n-1}$ are also shown semi-transparently.
	The parameter space becomes increasingly constrained as $n$ increases.
	}
\end{figure}

Now, we sequentially apply the linear map $\mathcal{E}$ to density matrices corresponding to points in the Bloch ball.
Then, the parameter spaces of the resulting density matrices are
\begin{equation}\label{eq:bn_def}
	\begin{split}
		\mathcal{B}_{n} &\coloneqq \{\mathbf{x}' \in \mathbb{R}^4 : \exists \mathbf{x} \in \mathcal{B}_{n-1} \text{ s.t. } \rho(\mathbf{x}') = \mathcal{E}(\rho(\mathbf{x}))\} \\
		&= \{\mathbf{x}' \in \mathbb{R}^4 : \exists \mathbf{x} \in \mathcal{B}_{0} \text{ s.t. } \rho(\mathbf{x}') = \mathcal{E}^n(\rho(\mathbf{x}))\}
	\end{split},
\end{equation}
where $n$ is a positive integer.
Given the geometrical description of the space of allowed state as a Blcoh ball, we can parameterize the output states more explicitly.
Notice the linear map $\mathcal{E}$ can be represented as an invertible matrix defined through $\mathcal{E}(\sigma_{\mu}) = \sum_{\nu=0}^{3} \sigma_{\nu}T_{\nu\mu}$ with $T \in \mathbb{M}_{4}(\mathbb{R})$.
Thanks to trace-preserving nature of $\mathcal{E}$, $T$ and $T^n$ are characterized by $\mathbf{a} \in \mathbb{R}^3$ and an invertible matrix $A$ as
\begin{align}
	\label{eq:T_CPTP}
	T = \begin{pmatrix}
		1 & \bm{0}^\transpose \\
		\mathbf{a} & A
	\end{pmatrix};\quad 
	T^n = \begin{pmatrix}
		1 & \bm{0}^\transpose \\
		\mathbf{b}_n & B_n
	\end{pmatrix},
\end{align}
where $\mathbf{b}_n = \sum_{k=0}^{n-1} A^k \mathbf{a}$ and $B_n = A^n$.
The transformation of density matrices, $\rho(\mathbf{x}') = \mathcal{E}^n(\rho(\mathbf{x}))$, implies
\begin{align}
	\begin{pmatrix}
		x'_0 \\
		\tilde{\mathbf{x}}'
	\end{pmatrix} = \begin{pmatrix}
		1 & \bm{0}^\transpose \\
		\mathbf{b}_n & B_n
	\end{pmatrix} \begin{pmatrix}
		x_0 \\
		\tilde{\mathbf{x}}
	\end{pmatrix};\  \begin{pmatrix}
		x_0 \\
		\tilde{\mathbf{x}}
	\end{pmatrix} \in \mathcal{B}_0.
\end{align}
Due to the normalization condition $x_0 = 1$, we have $x'_0 = 1$ and $\tilde{\mathbf{x}}' = B_n \tilde{\mathbf{x}} + \mathbf{b}_n$.
Since $\|\tilde{\mathbf{x}}\|_2 \leq x_0$, the output parameter space $\mathcal{B}_n$ is given explicitly by
\begin{equation}
	\begin{aligned}
		\mathcal{B}_n &= \{\mathbf{x} \in \mathbb{R}^4 : \|B_{n}^{-1}(\tilde{\mathbf{x}} - \mathbf{b}_n)\|_2 \leq x_0, x_0 = 1\} \\
		&\simeq \{\tilde{\mathbf{x}} \in \mathbb{R}^3 : \|B_{n}^{-1}(\tilde{\mathbf{x}} - \mathbf{b}_n)\|_2 \leq 1\}
	\end{aligned}\ ,
\end{equation}
which defines a nested set of ellipsoids contained inside the Bloch ball, as is demonstrated in Fig.~\ref{fig:bloch}.

\subsection{Linear matrix inequality and second-order cone}
While the qubit example demonstrates the shrinking of the space of propagated states as the same map is repeatedly applied, the qubit is special in that the Bloch ball, defined through a quadratic inequality, is an exact description of the space of feasible density matrices only in the qubit case.
To handle Hilbert space dimensions higher than two, we now return to the general matrix-type constraints $\mathcal{M}(\mathbf{x})$ and scalar-type constraints $f(\mathbf{x})$.
The matrix-valued function $\mathcal{M}$ is expressed as
\begin{align}
	\label{eq:LMI}
	\mathcal{M}(\mathbf{x}) = \sum_{\alpha=0}^{r-1} x_{\alpha} A_{\alpha},
\end{align}
where $\{A_{\alpha}\}_{\alpha=0}^{r-1}$ are linearly independent Hermitian matrices.
The inequality $\mathcal{M}(\mathbf{x}) \succeq 0$ is called a \textit{linear matrix inequality (LMI)}.
LMIs appear natural in various problems in quantum physics.
Examples include the positivity of physical density matrices $\rho(\mathbf{x}) \succeq 0$, separability criteria for bipartite quantum states~\cite{DPS_2004}, and $N$-representability conditions~\cite{QChem_boot2}.
In fact, $\mathcal{B}_n$ as defined in Eq.\ \eqref{eq:bn_def} can also be written as an LMI.
When we have LMIs, the problem~\eqref{eq:problem} is solved by \textit{semidefinite~programming~(SDP)}.

In contrast, the geometrical description surrounding the shrinking Bloch ball can be identified with scalar-type constraints. Here, we consider functions given by
\begin{equation}
    \label{eq:SOC}
    \begin{split}
        f(\mathbf{x}) = \mathbf{p}^\top \mathbf{x} + q - \|S \mathbf{x} + \mathbf{t}\|_2 \geq 0
    \end{split},
\end{equation}
where $\mathbf{p} \in \mathbb{R}^r$, $q \in \mathbb{R}, \mathbf{t} \in \mathbb{R}^{r-1}$, and $S$ is a $(r-1) \times r$-dimensional real-valued matrix.
In the single qubit example, we have $\mathbf{p} = (1,0,0,0)^\transpose, q=0, \mathbf{t} = -B_{n}^{-1}\mathbf{b}_n$, and $S = \begin{pmatrix}
    \mathbf{0} & B_{n}^{-1}
\end{pmatrix}$.
The set defined by Eq.~\eqref{eq:SOC}, i.e., $\{\mathbf{x} : f(\mathbf{x}) \geq 0\}$, is known as a \textit{second-order cone (SOC)}.
When only SOC-type constraints are present, problem~\eqref{eq:problem} can be solved via \textit{SOC~programming~(SOCP)}, which is computationally much less demanding than SDP.

SOC-type constraints also arise naturally as a relaxation of the positivity of density matrices: Let $\rho$ be a $\chi$-dimensional physical density matrix parameterized by $\rho = \sum_{\alpha=0}^{\chi^2-1} x_{\alpha} h_{\alpha}$, where $\{h_{\alpha}\}_{\alpha=0}^{\chi^2-1}$ is an orthonormal basis set of $\chi$-dimensional Hermitian matrices satisfying $\Tr[h_{\alpha}] = \chi\delta_{\alpha 0}$ and $\Tr[h_{\alpha}h_{\beta}] = \chi\delta_{\alpha \beta}$ for all $\alpha, \beta$.
Since $\rho$ is physical, it must be positive semidefinite, i.e., $\rho \succeq 0$.
We note that $\Tr[\rho^2] \leq \Tr[\rho]^2$ holds for $\rho \succeq 0$, which results in a form described by Eq.\ \eqref{eq:SOC}
\begin{equation}
	\label{eq:SOC_from_vanilla_PSD}
	\sqrt{\chi-1}x_0 - \sqrt{\sum_{\alpha=1}^{\chi^2-1} x_{\alpha}^2} \geq 0,
\end{equation}
where $x_0 = \Tr[\rho]/\chi > 0$.

\subsection{Bootstrap approach for contracting MPS}
Here, we demonstrate how the necessary conditions defining problem \eqref{eq:problem} can be obtained for translationally invariant MPSs.
An MPS is defined by a rank-3 tensor $M$ graphically represented as
\begin{equation}
	\label{eq:MPS_definition}
	\begin{split}
		M_{ij}^{s} = \begin{array}{c}
			\begin{tikzpicture}[scale=0.6,baseline={([yshift=-0.75ex] current bounding box.center)}]
				\begin{scope}[shift={(0,0.)}]
					\draw[thick, fill=tensorcolor] (-0.5,-0.5) rectangle (0.5,0.5);
					\node at (0,0) {$M$};
					\draw[thick] (0.5,0) -- (1,0);
					\node at (1.25,0) {$j$};
					\draw[thick] (-0.5,0) -- (-1,0);
					\node at (-1.25,0) {$i$};
					\draw[thick] (0,0.5) -- (0,1);
					\node at (0,1.25) {$s$};
				\end{scope};
			\end{tikzpicture}
		\end{array}
	\end{split} \in \mathbb{C}\ \ ,
\end{equation}
where $i, j = 0, \cdots \chi-1$ represent auxiliary degrees of freedom, and $s = 0, \cdots D-1$ denotes the physical degrees of freedom at each site.
The expectation value is evaluated via
\begin{align}
	\label{eq:expectation_value_MPS}
	\begin{array}{c}
		\begin{tikzpicture}[scale=0.6,baseline={([yshift=-0.75ex] current bounding box.center)}]
			\begin{scope}[shift={(1.5, -0.5)}]
				\draw[thick, fill=tensorcolor] (-0.5,-0.5) rectangle (0.5,0.5);
				\node at (0,0) (centerO) {$O$};
				\draw[thick] (0,-0.5) -- (0,-1);
			\end{scope}
			\begin{scope}[shift={(0,1.)}]
				\draw[thick, fill=tensorcolor] (-0.5,-0.5) rectangle (0.5,0.5);
				\node at (0,0) {$M$};
				\draw[thick] (0.5,0) -- (1,0);
				\draw[thick] (-0.5,0) -- (-1,0);
				\draw[thick] (0,-0.5) -- (0,-2.5);
			\end{scope}
			\begin{scope}[shift={(1.5,1.)}]
				\draw[thick, fill=tensorcolor] (-0.5,-0.5) rectangle (0.5,0.5);
				\node at (0,0) {$M$};
				\draw[thick] (0.5,0) -- (1,0);
				\draw[thick] (-0.5,0) -- (-1,0);
				\draw[thick] (0,-0.5) -- (0,-1);
			\end{scope}
			\begin{scope}[shift={(3.,1.)}]
				\draw[thick, fill=tensorcolor] (-0.5,-0.5) rectangle (0.5,0.5);
				\node at (0,0) {$M$};
				\draw[thick] (0.5,0) -- (1,0);
				\draw[thick] (-0.5,0) -- (-1,0);
				\draw[thick] (0,-0.5) -- (0,-2.5);
			\end{scope}
			\begin{scope}[shift={(6.,1.)}]
				\draw[thick, fill=tensorcolor] (-0.5,-0.5) rectangle (0.5,0.5);
				\node at (0,0) {$M$};
				\draw[thick] (0.5,0) -- (1,0);
				\draw[thick] (-0.5,0) -- (-1,0);
				\draw[thick] (0,-0.5) -- (0,-2.5);
				\draw[thick, dotted] (-1.,0) -- (-2.,0);
			\end{scope}
			\begin{scope}[shift={(-3.,1.)}]
				\draw[thick, fill=tensorcolor] (-0.5,-0.5) rectangle (0.5,0.5);
				\node at (0,0) {$M$};
				\draw[thick] (0.5,0) -- (1,0);
				\draw[thick] (-0.5,0) -- (-1,0);
				\draw[thick] (0,-0.5) -- (0,-2.5);
				\draw[thick, dotted] (1.,0) -- (2.,0);
			\end{scope}
			\begin{scope}[shift={(-3,-2.)}]
				\draw[thick, fill=tensorcolor] (-0.5,-0.5) rectangle (0.5,0.5);
				\node at (0,0) {$\bar{M}$};
				\draw[thick] (0.5,0) -- (1,0);
				\draw[thick] (-0.5,0) -- (-1,0);
				\draw[thick, dotted] (1.,0) -- (2.,0);
			\end{scope}
			\begin{scope}[shift={(0,-2.)}]
				\draw[thick, fill=tensorcolor] (-0.5,-0.5) rectangle (0.5,0.5);
				\node at (0,0) {$\bar{M}$};
				\draw[thick] (0.5,0) -- (1,0);
				\draw[thick] (-0.5,0) -- (-1,0);
			\end{scope}
			\begin{scope}[shift={(1.5,-2.)}]
				\draw[thick, fill=tensorcolor] (-0.5,-0.5) rectangle (0.5,0.5);
				\node at (0,0) {$\bar{M}$};
				\draw[thick] (0.5,0) -- (1,0);
				\draw[thick] (-0.5,0) -- (-1,0);
			\end{scope}
			\begin{scope}[shift={(3.,-2.)}]
				\draw[thick, fill=tensorcolor] (-0.5,-0.5) rectangle (0.5,0.5);
				\node at (0,0) {$\bar{M}$};
				\draw[thick] (0.5,0) -- (1,0);
				\draw[thick] (-0.5,0) -- (-1,0);
			\end{scope}
			\begin{scope}[shift={(6,-2.)}]
				\draw[thick, fill=tensorcolor] (-0.5,-0.5) rectangle (0.5,0.5);
				\node at (0,0) {$\bar{M}$};
				\draw[thick] (0.5,0) -- (1,0);
				\draw[thick] (-0.5,0) -- (-1,0);
				\draw[thick, dotted] (-1.,0) -- (-2.,0);
			\end{scope}
			\begin{scope}[decoration={brace, amplitude=10pt}, thick]
				\draw[decorate] (-3.7, 1.8) -- (1.0, 1.8)
					node [midway, above=12pt, font=\small] {$L$};
				\draw[decorate] (2.0, 1.8) -- (6.7, 1.8)
					node [midway, above=12pt, font=\small] {$R$};
			\end{scope}
		\end{tikzpicture}
		\end{array}
\end{align}
with a proper boundary condition and the normalization of the MPS.

We now construct scalar-type and matrix-type constraints associated with a given MPS.
The linear map $\calE$ is a transfer matrix defined by
\begin{equation}
	\label{eq:transfer_matrix}
	\calE(X) \coloneqq \sum_{s=0}^{D-1} M^{s} X [M^{s}]^{\dagger} = \begin{array}{c}
		\begin{tikzpicture}[scale=0.6,baseline={([yshift=-0.75ex] current bounding box.center)}]
			\begin{scope}[shift={(0,0)}]
				\draw[thick, fill=tensorcolor] (-0.5,-0.5) rectangle (0.5,0.5);
				\node at (0,0) {$M$};
				\draw[thick] (0.5,0) -- (1,0);
				\draw[thick] (-0.5,0) -- (-1,0);
				\draw[thick] (0,-0.5) -- (0,-1);
			\end{scope};
			\begin{scope}[shift={(0,-1.5)}]
				\draw[thick, fill=tensorcolor] (-0.5,-0.5) rectangle (0.5,0.5);
				\node at (0,0) {$\bar{M}$};
				\draw[thick] (0.5,0) -- (1,0);
				\draw[thick] (-0.5,0) -- (-1,0);
			\end{scope};
			\begin{scope}[shift={(1., -0.75)}]
				\draw[thick, fill=tensorcolor] (0, 0.) circle (0.4);
				\node at (0, 0.) {$X$};
				\draw[thick] (0.,0.4) -- (0,0.75);
				\draw[thick] (0.,-0.4) -- (0,-0.75);
			\end{scope}
		\end{tikzpicture}
	\end{array}\ \ .
\end{equation}
Then, we represent $\calE$ by a $\chi^{2}$-dimensional matrix $T$ such that
\begin{align}
	\calE(h_{\alpha}) = \sum_{\beta=0}^{\chi^2-1} h_{\beta} T_{\beta\alpha};\quad T_{\beta\alpha} = \Tr[h_{\beta} \calE(h_{\alpha})]/\chi .
\end{align}
For simplicity, we assume that $\mathrm{rank}(T) = \chi^2$.
It is straightforward to extend the following discussions to the case of $\mathrm{rank}(T) < \chi^2$. See Supplementary Materials~\cite{SM} for details.
As discussed in the general framework, rather than specifying a boundary environment at the right boundary, we require a positivity constraint for environments at the right boundary.
Starting from such a vanilla constraint, we propagate it to the target site.

\subsubsection{MPS in canonical form}
\begin{figure}[t]
	\centering
	\includegraphics[width=0.99\columnwidth]{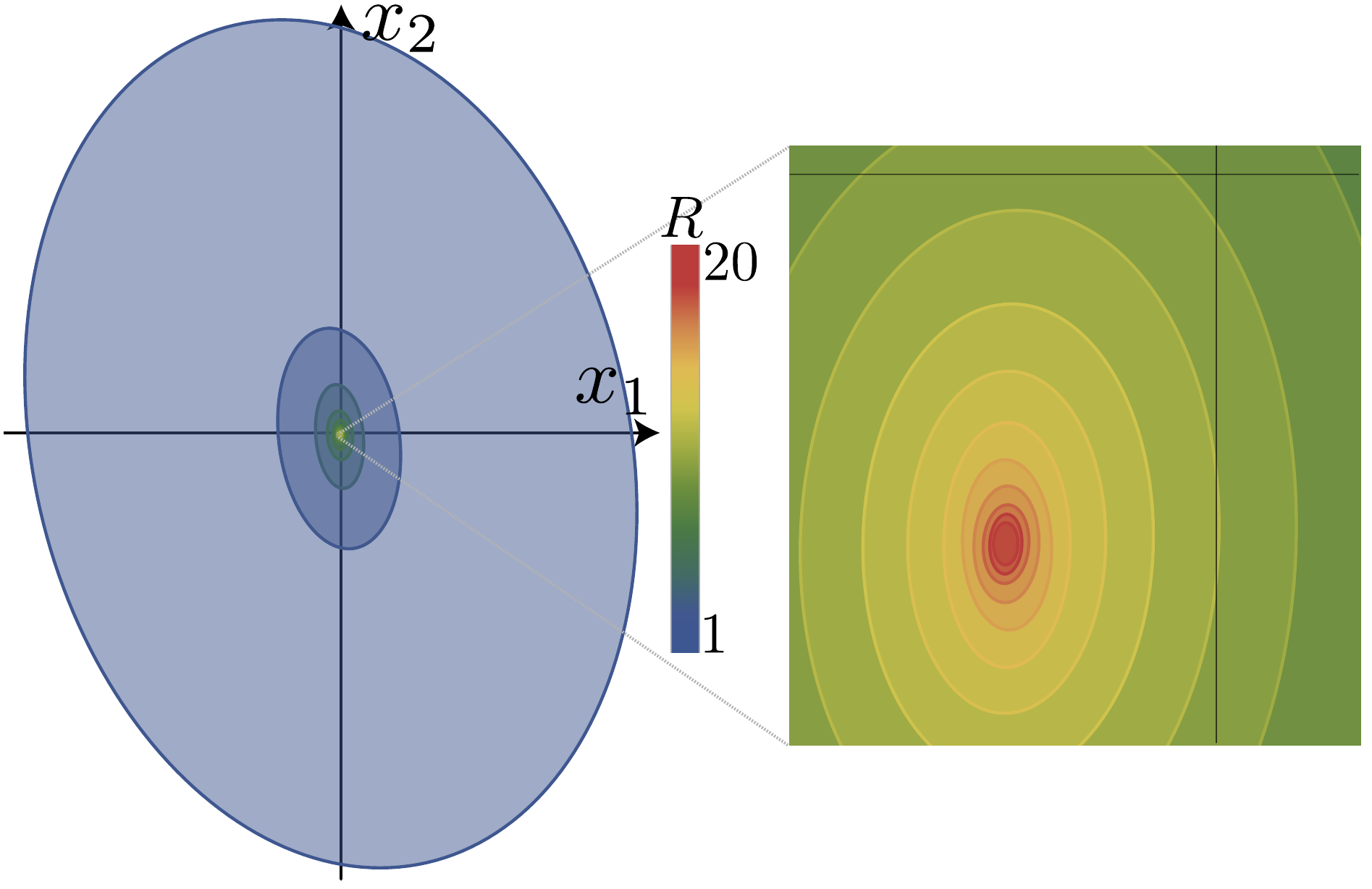}
	\caption{\label{fig:region_plot_ellipsoid}
	Projected parameter space of environments at the target site for a random MPS with $\chi = 20$.
	The full parameter space is defined by $\mathcal{C}_{\Lambda_{R}}\ (1 \leq R \leq 20)$ and the normalization slice $x_0 = 1/\chi$, whose intersection forms a hyper-ellipsoid.
	For visualization purposes, the parameter space is projected onto the $x_1$-$x_2$ plane.
	The right panel shows an enlarged view of the region $-1.5 \times 10^{-4} \leq x_1 \leq 5 \times 10^{-5}$ and $-2 \times 10^{-4} \leq x_2 \leq 1 \times 10^{-5}$.
	Similar tightening can be observed for other projections of the parameter space.
	}
\end{figure}

As a first demonstration, we show SOC-type constraints can provide tight bounds on the MPS contraction results when the MPS is first brought into a canonical form. Suppose the MPS is in the left-canonical form
\begin{equation}
	\label{eq:left-canonical}
	\begin{aligned}
		\begin{array}{c}
			\begin{tikzpicture}[scale=0.6,baseline={([yshift=-0.75ex] current bounding box.center)}]
				\begin{scope}[shift={(0,0)}]
					\draw[thick, fill=tensorcolor] (-0.5,-0.5) rectangle (0.5,0.5);
					\node at (0,0) {$M$};
					\draw[thick] (0.5,0) -- (1,0);
					\draw[thick] (-0.5,0) -- (-1,0);
					\draw[thick] (0,-0.5) -- (0,-1);
				\end{scope};
				\begin{scope}[shift={(0,-1.5)}]
					\draw[thick, fill=tensorcolor] (-0.5,-0.5) rectangle (0.5,0.5);
					\node at (0,0) {$\bar{M}$};
					\draw[thick] (0.5,0) -- (1,0);
					\draw[thick] (-0.5,0) -- (-1,0);
					\draw[thick] (-1.,0) -- (-1.,1.5);
				\end{scope};
			\end{tikzpicture}
		\end{array}
		= \begin{array}{c}
			\begin{tikzpicture}[scale=0.6,baseline={([yshift=-0.75ex] current bounding box.center)}]
				\begin{scope}[shift={(0,0)}]
					\draw[thick] (0,-1.) -- (0,1);
				\end{scope};
			\end{tikzpicture}
		\end{array}
	\end{aligned}.
\end{equation}
Taking the semi-infinite limit $L \to \infty$ while keeping $R$ finite in Eq.~\eqref{eq:expectation_value_MPS}, we focus on constraining the parameter space of the right environments $E_R(\mathbf{x}) = \sum_{\alpha=0}^{\chi^{2}-1} x_{\alpha} h_{\alpha}$, parametrized by $\mathbf{x} \in \mathbb{R}^{\chi^2}$. The assumption of the left-canonical form simplifies the normalization condition to $\Tr[E_R(\mathbf{x})] = 1$, which implies $x_0 = 1/\chi$.
As such, the transfer matrix is a CPTP map and the problem becomes a natural higher-dimensional generalization to the qubit example we studied before, with $T$ taking a similar form to Eq.~\eqref{eq:T_CPTP}.
We require the right boundary environments to be positive semidefinite, and further relax the positive semidefiniteness to SOC-type constraints, as seen in Eq.~\eqref{eq:SOC_from_vanilla_PSD}.
As such, the parameter space of boundary environments is given by
\begin{align}
	\mathcal{C}_{\Lambda_0} \coloneqq \left\{\mathbf{x} = \begin{pmatrix}
		x_0 \\
		\tilde{\mathbf{x}}
	\end{pmatrix} \in \mathbb{R}^{\chi^2}: \|\tilde{\mathbf{x}}\|_2 \leq \sqrt{\chi-1}x_0,\ x_0 = 1/\chi\right\},
\end{align}
Since both $\mathcal{C}_{\Lambda_0}$ and $T$ have the same structural form as in the qubit example, we can follow the same procedure to characterize the parameter space of the environment at the target site.
The resulting parameter space after $R$ propagations, denoted by $\mathcal{C}_{\Lambda_{R}}$, is a hyper-ellipsoid.
Figure~\ref{fig:region_plot_ellipsoid} shows the parameter space $\mathcal{C}_{\Lambda_{R}}\ (0 \leq R \leq 20)$ for a random MPS with $\chi = 20$, which shows that a tight bound on the right environment state, and hence any local observables at the target site, can be obtained from the propagation of the SOC-type constraints.

\subsubsection{MPS without canonical form}
Next, we lift the assumption on the canonical form of the MPS. This implies the transfer matrix no longer defines a trace-preserving map in general. Nevertheless, we now show that tight bounds on the environment state could still be obtained from LMI constraints.
We again start with the vanilla positivity at the right boundary:
\begin{equation}
	\mathcal{S}_0 \coloneqq \left\{\mathbf{x} \in \mathbb{R}^{\chi^2}: \mathcal{M}^{(0)}(\mathbf{x}) = \sum_{\alpha=0}^{\chi^{2}-1} x_{\alpha} h_{\alpha} \succeq 0\right\},
\end{equation}
in which matrices $\{A^{(0)}_{\alpha}\}_{\alpha=0}^{\chi^{2}-1}$ in Eq.~\eqref{eq:LMI} correspond to $A^{(0)}_{\alpha} = h_{\alpha}$.
It is convenient to express Hermitian matrices defining an LMI as a linear combination of an orthonormal basis set $\{h_{\alpha}\}_{\alpha}$, such that $A_{\alpha} = \sum_{\beta} h_{\beta}\mathcal{L}_{\beta\alpha}$.
In this representation, $\mathcal{L}$ completely encodes information about the LMI.
For $\{A_{\alpha}\}_{\alpha}$, the corresponding coefficient matrix $\mathcal{L}^{(0)}$ is simply an identity matrix.
Similar to the case of SOC-type constraints, we systematically propagate these constraints to the target site.
The coefficient matrix $\mathcal{L}$ can be shown to be propagated by $\mathcal{L}^{(i+1)} = \mathcal{L}^{(i)}T^{-1}$~\cite{SM}.
Then, we construct the parameter space of right environments at the target site:
\begin{equation}
	\mathcal{S}_{R} \coloneqq \left\{\mathbf{x} \in \mathbb{R}^{\chi^2}: \mathcal{M}^{(R)}(\mathbf{x}) = \sum_{\alpha=0}^{\chi^{2}-1} x_{\alpha} A^{(R)}_{\alpha} \succeq 0\right\},
\end{equation}
where $A^{(R)}_{\alpha} = \sum_{\beta} h_{\beta}\mathcal{L}^{(R)}_{\beta\alpha}$ with $\mathcal{L}^{(R)} = T^{-R}$.

\begin{figure}[t]
	\centering
	\includegraphics[width=\columnwidth]{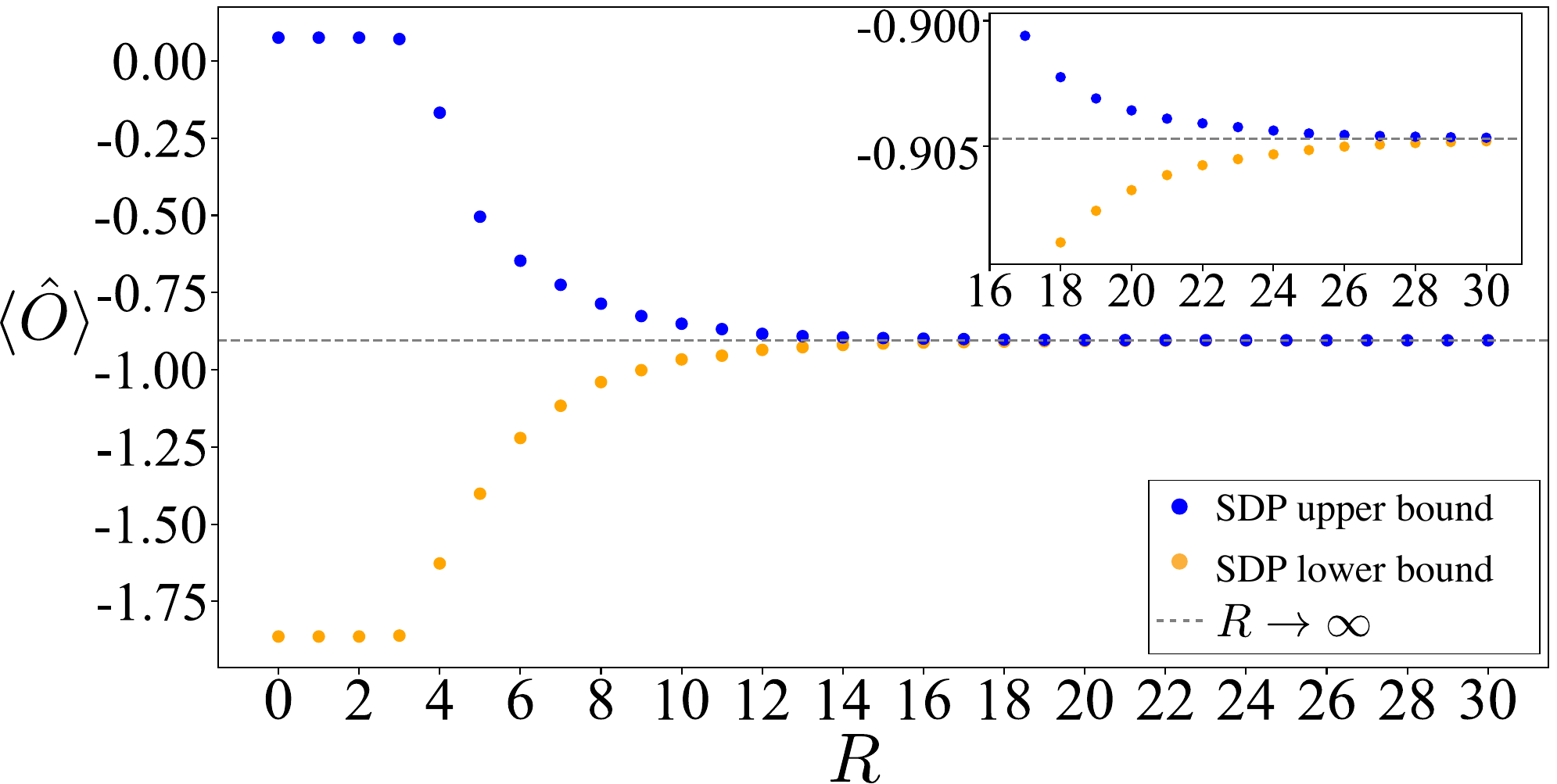}
	\caption{\label{fig:generic_result}
	Numerical bounds on the expectation value of a randomly chosen observable for a randomly generated MPS with bond dimension $\chi=8$.
	Here, we take $L=R$ and compute the bounds for $0\leq R \leq 30$.
	The result for $R \to \infty$ is computed by left- and right-fixed point environments.
	}
\end{figure}

We now lift the assumption of the semi-infinite limit $L \to \infty$.
For finite $L$, by employing the same procedures, we can construct the parameter space $\mathcal{S}_{L}$ of left environments $E_{L}$.
Then, the left- and right-environments, $E_{L}(\mathbf{x}^{(L)})$ and $E_{R}(\mathbf{x}^{(R)})$, are parameterized by $\mathbf{x}^{(L)} \in \mathcal{S}_{L}$ and $\mathbf{x}^{(R)} \in \mathcal{S}_{R}$, respectively.
Once we have left- and right-environments, we can evaluate the expectation value of the observable $O$ via
\begin{equation}
	\begin{aligned}
		\begin{array}{c}
			\begin{tikzpicture}[scale=0.6,baseline={([yshift=-0.75ex] current bounding box.center)}]
				\begin{scope}[shift={(0,1.)}]
					\draw[thick, fill=tensorcolor] (-0.5,-0.5) rectangle (0.5,0.5);
					\node at (0,0) {$M$};
					\draw[thick] (0.5,0) -- (1,0);
					\draw[thick] (-0.5,0) -- (-1,0);
					\draw[thick] (0,-0.5) -- (0,-1);
				\end{scope};
				\begin{scope}[shift={(0, -0.5)}]
					\draw[thick, fill=tensorcolor] (-0.5,-0.5) rectangle (0.5,0.5);
					\node at (0,0) {$O$};
				\end{scope};
				\begin{scope}[shift={(0,-2.)}]
					\draw[thick, fill=tensorcolor] (-0.5,-0.5) rectangle (0.5,0.5);
					\node at (0,0) {$\bar{M}$};
					\draw[thick] (0.,0.5) -- (0,1.);
					\draw[thick] (0.5,0) -- (1,0);
					\draw[thick] (-0.5,0) -- (-1,0);
				\end{scope};
				\begin{scope}[shift={(1.75,-0.5)}]
					\draw[thick, fill=tensorcolor] (-0.75,-1.) rectangle (0.75,1.);
					\node at (0,0) {$E_R$};
					\draw[thick] (0.,1.) -- (0,1.5);
					\draw[thick] (0.,-1.) -- (0,-1.5);
					\draw[thick] (-0.75,1.5) -- (0.,1.5);
					\draw[thick] (-0.75,-1.5) -- (0.,-1.5);
				\end{scope}
				\begin{scope}[shift={(-1.75,-0.5)}]
					\draw[thick, fill=tensorcolor] (-0.75,-1.) rectangle (0.75,1.);
					\node at (0,0) {$E_L$};
					\draw[thick] (0.,1.) -- (0,1.5);
					\draw[thick] (0.,-1.) -- (0,-1.5);
					\draw[thick] (0.75,1.5) -- (0.,1.5);
					\draw[thick] (0.75,-1.5) -- (0.,-1.5);
				\end{scope}
			\end{tikzpicture}
		\end{array}
	\end{aligned}\ \ .
\end{equation}
However, this objective function is bilinear with respect to $\mathbf{x}^{(L)}$ and $\mathbf{x}^{(R)}$, rendering the corresponding optimization problem non-convex.
To circumvent this difficulty, we introduce a further relaxation. Physically, left- and right-environments are completely disentangled, meaning the total environment is a product state, $E = E_L\otimes E_R$.
We relax this product structure by allowing the total environment to be a potentially entangled bipartite state, provided it remains positive semidefinite, i.e., $E \succeq 0$, and its reduced operators on the left and right subsystems satisfy local conditions analogous to the quantum marginal problem:
\begin{equation}
	\label{eq:conditions_for_total_environment}
	\begin{aligned}
		&\Tr_{R}[E(\mathbf{x})] \in \{E_{L}(\mathbf{x}^{(L)}): \mathbf{x}^{(L)} \in \mathcal{S}_{L}\}\\
		&\Tr_{L}[E(\mathbf{x})] \in \{E_{R}(\mathbf{x}^{(R)}) : \mathbf{x}^{(R)} \in \mathcal{S}_{R}\}
	\end{aligned}\ .
\end{equation}
To further tighten this relaxation, we impose an additional constraint on the total environments, which originates from the tensor product of the two parameter spaces:
\begin{align}
	\left\{\mathbf{x} \in \mathbb{R}^{\chi^4}: \sum_{\alpha, \beta}x_{(\alpha,\beta)}A_{\alpha}^{(L)}\otimes A_{\beta}^{(R)} \succeq 0\right\}.
\end{align}
Such a relaxation enables us to formulate an SDP to optimize the following objective function with the normalization condition:
\begin{equation}
	\begin{aligned}
		\begin{array}{c}
			\begin{tikzpicture}[scale=0.6,baseline={([yshift=-0.75ex] current bounding box.center)}]
				\begin{scope}[shift={(0,1.)}]
					\draw[thick, fill=tensorcolor] (-0.5,-0.5) rectangle (0.5,0.5);
					\node at (0,0) {$M$};
					\draw[thick] (0.,-0.5) -- (0,-1.);
					\draw[thick] (-0.25,0.5) -- (-0.25,0.75);
					\draw[thick] (-0.25,0.75) -- (1.5,0.75);
					\draw[thick] (1.5,0.75) -- (1.5,-0.5);
					\draw[thick] (0.25,0.5) -- (0.25,1.);
					\draw[thick] (0.25,1.) -- (2.,1.);
					\draw[thick] (2.,1.) -- (2.,-0.5);
				\end{scope};
				\begin{scope}[shift={(0, -0.5)}]
					\draw[thick, fill=tensorcolor] (-0.5,-0.5) rectangle (0.5,0.5);
					\node at (0,0) {$O$};
				\end{scope};
				\begin{scope}[shift={(0,-2.)}]
					\draw[thick, fill=tensorcolor] (-0.5,-0.5) rectangle (0.5,0.5);
					\node at (0,0) {$\bar{M}$};
					\draw[thick] (0.,0.5) -- (0,1.);
					\draw[thick] (-0.25,-0.5) -- (-0.25,-0.75);
					\draw[thick] (-0.25,-0.75) -- (1.5,-0.75);
					\draw[thick] (1.5,-0.75) -- (1.5, 1.5);
					\draw[thick] (0.25,-0.5) -- (0.25,-1.);
					\draw[thick] (0.25,-1.) -- (2.,-1.);
					\draw[thick] (2.,-1.) -- (2., 1.5);
				\end{scope};
				\begin{scope}[shift={(1.75,-0.5)}]
					\draw[thick, fill=tensorcolor] (-0.75,-1.) rectangle (0.75,1.);
					\node at (0,0) {$E(\mathbf{x})$};
				\end{scope}
			\end{tikzpicture}
		\end{array}
		\ \text{ with }\
		\begin{array}{c}
		\begin{tikzpicture}[scale=0.6,baseline={([yshift=-0.75ex] current bounding box.center)}]
			\begin{scope}[shift={(0,0.5)}]
				\draw[thick, fill=tensorcolor] (-0.5,-0.5) rectangle (0.5,0.5);
				\node at (0,0) {$M$};
				\draw[thick] (0.,-0.5) -- (0,-1.);
				\draw[thick] (-0.25,0.5) -- (-0.25,0.75);
				\draw[thick] (-0.25,0.75) -- (1.5,0.75);
				\draw[thick] (1.5,0.75) -- (1.5,-0.5);
				\draw[thick] (0.25,0.5) -- (0.25,1.);
				\draw[thick] (0.25,1.) -- (2.,1.);
				\draw[thick] (2.,1.) -- (2.,-0.5);
			\end{scope};
			\begin{scope}[shift={(0,-1.5)}]
				\draw[thick, fill=tensorcolor] (-0.5,-0.5) rectangle (0.5,0.5);
				\node at (0,0) {$\bar{M}$};
				\draw[thick] (0.,0.5) -- (0,1.);
				\draw[thick] (-0.25,-0.5) -- (-0.25,-0.75);
				\draw[thick] (-0.25,-0.75) -- (1.5,-0.75);
				\draw[thick] (1.5,-0.75) -- (1.5, 1.5);
				\draw[thick] (0.25,-0.5) -- (0.25,-1.);
				\draw[thick] (0.25,-1.) -- (2.,-1.);
				\draw[thick] (2.,-1.) -- (2., 1.5);
			\end{scope};
			\begin{scope}[shift={(1.75,-0.5)}]
				\draw[thick, fill=tensorcolor] (-0.75,-1.) rectangle (0.75,1.);
				\node at (0,0) {$E(\mathbf{x})$};
			\end{scope}
		\end{tikzpicture}
	\end{array} = 1
	\end{aligned}\ \ .
\end{equation}
Figure~\ref{fig:generic_result} shows the numerical results for a random non-canonical MPS with $\chi = 8$.
One can see that the bounds are tight for large $R$ and $L$.

\section{Discussion}
In this work, we proposed a rigorous approach for computing two-sided bounds on expectation values of local observables in tensor networks.
After introducing our general concept of bootstrapping tensor-network contractions, we first presented a concrete algorithm to derive physical constraints on MPS environments at a target site.
We then demonstrated numerically how tight these bounds can be for a given MPS and observable.

As a remark, our approach does not yield tight bounds for all MPSs.
A prime example is the GHZ state~\cite{GHZ}.
While our method provides tight bounds for the Pauli-$X$ and Pauli-$Y$ operators, it fails to do so for the Pauli-$Z$ operator.
This is because the corresponding transfer matrix is not irreducible for the GHZ state.
This behavior is consistent with the fact that the expectation value of the Pauli-$Z$ operator inherently depends on the choice of fixed-point environments in the thermodynamic limit and is therefore not well-defined without committing to a specific choice.

Lastly, while the formulation of tensor-network contraction as an optimization problem amenable to bootstrap relaxation is general, the systematic construction of useful relaxations beyond the MPS problem we studied remains unsolved.
In two dimensions, the linear map in Fig.~\ref{fig:ipeps}(d) can be decomposed into smaller linear maps, which define several convex sets for the center-site environments.
Similar to MPSs, the total environment $E$ must satisfy various marginal-problem-like conditions, leading to an SDP relaxation and bounds on observable expectation values. However, there is a general tug-of-war between the computational feasibility of the resulting SDP problem and the tightness of the bounds coming from the relaxation. 
The development of a practical algorithm extending the developed approach to higher dimensions therefore remains an important open question.

\begin{acknowledgments}
	We thank Eslam Khalaf and Patrick Ledwith for useful discussions.
	S.O.~was supported by RIKEN Special Postdoctoral Researchers Program, RIKEN Quantum, KAKENHI Grant No.~JP25K17322 from the Japan Society for the Promotion of Science.
	H.C.P. and Y.Z.~were supported by the Hong Kong Research Grants Council through grants C7037-22G-1 and CRS\_CUHK401/22, and the Croucher Foundation through CIA23SC01.
	S.O.~and H.C.P.~thank the Yukawa Institute for Theoretical Physics at Kyoto University for fruitful discussions during the YITP workshop  YITP-I-25-02 on ``Recent Developments and Challenges in Tensor Networks: Algorithms, Applications to Science, and Rigorous Theories.''

	Our numerical simulations were performed using the open-source software package PICOS~\cite{picos}.
	The authors disclose that generative AI tools were used for improving English expressions.

\end{acknowledgments}

\bibliography{ref}

\end{document}